# Implementation of Security Systems for Detection and Prevention of Data Loss/Leakage at Organization via Traffic Inspection

Mir Hassan[1], Chen Jincai[1], Adnan Iftekhar[2], Adnan Shehzad[3], Xiaohui Cui[2]

[1]Wuhan National Laboratory for Optoelectronics, Huazhong University of Science and Technology, Wuhan, China

[2]Key Laboratory of Aerospace Information Security and Trusted Computing, Ministry of Education, Wuhan University, Wuhan 430000, China

[3]Department of Computer Science, Maharishi International University, Fairfield, Iowa, United States

hassanmir@hust.edu.cn; jcchen@hust.edu.cn; adnan@whu.edu.cn; ashehzad@mum.edu; xcui@whu.edu.cn;

*Correspondence should be addressed to jcchen@hust.edu.cn

## Abstract

Data Loss/Leakage Prevention (DLP) continues to be the main issue for many large organizations. There are multiple numbers of emerging security attach scenarios and a limitless number of overcoming solutions. Today's enterprises' major concern is to protect confidential information because a leakage that compromises confidential data means that sensitive information is in competitors' hands. Different data types need to be protected. However, our research is focused only on data in motion (DIM) i-e data transferred through the network. The research and scenarios in this paper demonstrate a recent survey on information and data leakage incidents, which reveals its importance and also proposed a model solution that will offer the combination of previous methodologies with a new way of pattern matching by advanced content checker based on the use of machine learning to protect data within an organization and then take actions accordingly. This paper also proposed a DLP deployment design on the gateway level that shows how data is moving through intermediate channels before reaching the final destination using the squid proxy server and ICAP server.

## 1. Introduction

Data loss hindrance might be a security threat that is quite distinctive following ancient classical security layers of protection. Over the last few decades, enterprises became progressively dependent on digital information to fulfill business goals. A vital amount of data processes involve parties each within and outside of organization network boundaries on any given business day. There are a few strategies for this data to travel. They impart a couple of models in a few structures, i-e, email messages, Information handling records, spreadsheets, database documents, and moment electronic correspondence. Quite a bit of this data is not delicate in any case; as a rule, it is ordered as "Touchy or Exclusive," showing that this information should be shielded from unapproved access or presentation. This need can be driven by Data Loss/Leakage programs, which protect the data within the organization by defining some rules and policies and monitoring every kind of data going outside as well as inside the organization through several means.

Detecting and forestalling information loss can defend against complete harm, competitive damage, and legal transactions. The DLP program is the mechanism by which an organization identifies their most sensitive data, where the data is authorized to be stored or processed, who or what applications should have access to the data, and how to protect from the loss of the sensitive data.[1]

Data leakage incidents have a great impact on the enterprise position [2]. According to IBM's survey report, 46 % of the companies experienced reputational harm because of information leakage and confidential employee information and member information. Data leakage incidents typically occur when employees performed usual daily life tasks, such as sending emails that contain sensitive information [3]. Concerns over this need make us compel to control better and protect sensitive information.

The next section comprises the DLP components that make DLP strategies and how they are involved in the DLP program. Data types, Data classification, and a threat to the actors are several components. However, there is a need to define, re-evaluate, and evolve these components during the DLP program's lifecycle.

## 1.1 Data Types.

Usually, the information is kept in two different ways, structured and unstructured. The kind of method utilizing the information elaborates the kind of information.

Three different data states need to be secured to achieve the organization's security goals, including data at rest, data in motion, and data in use. However, in this paper, we addressed data in motion only by a set of technologies provided by our DLP solution.

- Data at rest____ Information that is situated in document frameworks, databases, and other storage techniques.
- Data in motion____ Any information that is traveling through the system to the outside using system traffic.
- Data in use____ Information is use or information at the endpoints of the system (for example, information on gadgets, for example, USB, outside gadgets, PCs, and cell phones.) [4][5]

## 1.2 Data Defining and Classification

There are two main benefits of defining the data. First of all, defining data type will let the organization know where the sensitive data resides.  Secondly, it will let the organization know the method necessary for the classification of data types. Moreover, we also realized that either the data is structured or unstructured, as it is difficult to handle unstructured data.

In classification, Enterprises characterize the data attributes to ensure that the DLP program identifies the data according to policies defined. Sensitive data allocated types help organizations detect capabilities and avoid violating the data rules by generating a proper alert system.

Classifying the data is important for the DLP program. As the data is constantly changing its location, user, and type, there is a need for constant classification re-evaluation and policies to be defined within an organization. Standard built-in policies can reduce time to value and are less complex to implement [6].

## 1.3 DLP Threat Actors:

DLP was originally designed to alert organizations to the unintended misuse of internal data by an organizational employee, identifying broken business processes during the discovery [7].

There are two types of Threat Actors. There is a chance that sensitive information might be leaked, i-e Malicious Insider, and Malicious Outsider.

A malicious Insider threat is a member of an organization who has intentions to breach the administrative policies. The reason for this act may differ depending upon individuals; the employee could be leaving, or the employee could be a spy of the competitor organization and have been paid to keep the data.

Although Malicious outsider is not part of the organization, they have the same intentions as the malicious insider to break the rules and steal the data to harm organizational policies. Instances of this incorporate the endeavor to keep Sony pictures from discharging a disliked film by specific gatherings [8]. These styles of assaults are described as hacktivist vigilantism [9].

AIIM look into from the Business Watch Report titled "Data Privacy"– living by new standards" demonstrates that over half of respondents feel that half of the data theft occurs by internal staff or ex-staff, which shows the importance of data leakage prevention.

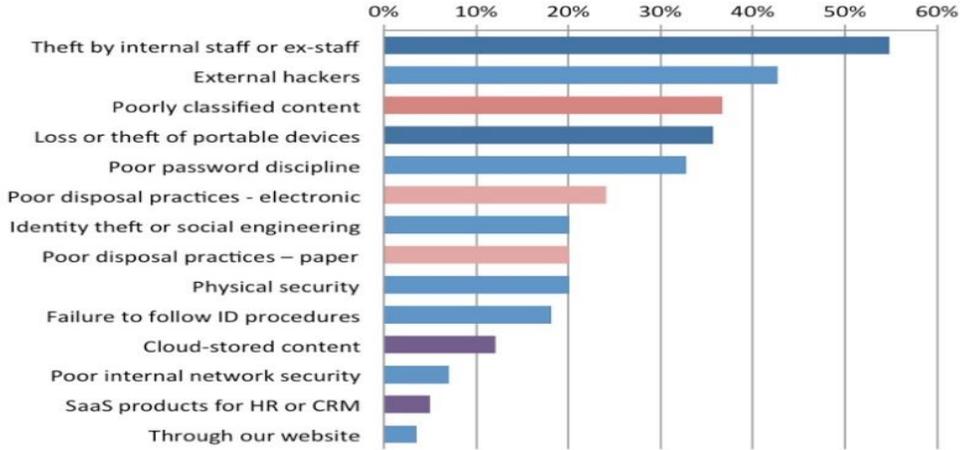

Figure 1: A study by AIIM showing that the organization is most vulnerable to data theft, loss, or exposure.

## 2. Literature Review

Various firms have currently started facilitating data and information leakage prevention solutions. Whereas some solutions protect 'data at rest' by limiting access to that and encrypting it, state of the art depends on strong and steady policies and pattern matching algorithms for information leak detection. Alternatively, related work in information leak prevention targets creating policies [11], building watermarking schemes [12], and distinguishing the forensic proof for post-mortem investigation [13].

To implementing a user-level policy language, hardware-implemented policies [11] are planned to ensure that the precise information does not arrive at the incredible output authorities through a network connection, files, and shared memory. The expected security system accredits pre-defined markers to the data. Rules are being implemented on the hardware level to assure the information stream abide by the policies. This solution's drawback associates the labor accelerated task of describing labels and policies and requires expensive hardware that supports information flow security.

Lee et al. [12] introduced data information leakage counteractive action from a forensics perspective. They perceived the arrangement of records required to find data spills on a working framework utilizing windows. The creators talk about that deferring the accumulation of the criminological information will have conspicuous impacts on the information leakage avoidance framework's productivity. Subsequently, they give an efficient method to assemble the central data required to discover information spills by working on five vital framework documents: installation record file, system event log, windows registry, browser history, and the core file in NTFS. Their methodology is limited to file system level information leaks on Windows platforms.

The artificial decoy theme of White et al. [13] elaborate the information leaks on massive databases of non-public records and proposes realistic decoy records to spot the source of information leaks, significantly when multiple databases are involved. By making unambiguously placeable, however logical individual records, the database can be digitally watermarked. Thus, an information leak from the database will contain the watermarks distinctive to the database in question hence declaring the leak's source. By nature, such a technique targets the post-mortem identification of the data leak source. The leak itself is real-time detection. However, a limitation to this approach is that an attacker can delete the watermark by any means.

L. Schiff et al.[14] develops the privacy-preserving schemes keeping in mind the privacy-preserving perceptive. Today it is difficult to believe that users or employees have to blindly trust the administrator or DLP engineer managing the intrusion detection or prevention system. This paper investigates this presumption. More precisely, they showed that it truly is feasible to decrease trust assumptions within the organizational network and proposed an intrusion detection system that is solely privacy conserving concerning user traffic and the principles utilized in IDS/IPS. The proposed PRI system enforced a single secured server; no hardware modifications are needed at the user end.

The current modern approach in information leak prevention targets matching different patterns that suffer from the overall deficiency of exploitation detection techniques: DLP engineer or administrator must outline the signatures. By Information leaks definition, signatures should be defined as per corporation basis, creating the widespread deployment of current information leak prevention tools a challenge. Besides that, the related work on data leak prevention and data mining relies on a forensics approach and primarily demonstrates post-mortem recognition

## 2.1 Survey of Data and Information Leakage Incidents

Regarding DLP, many threats exist, which leads to information and data exposure events. To improving the security system and forbid information loss/leakage incidents, the real goal is to understand and analyze the past incidents and attacks to take countermeasures.

This work makes use of information referred to as Datalossdb [15] and a report from Risk Based Security (RBS) [3] that provides

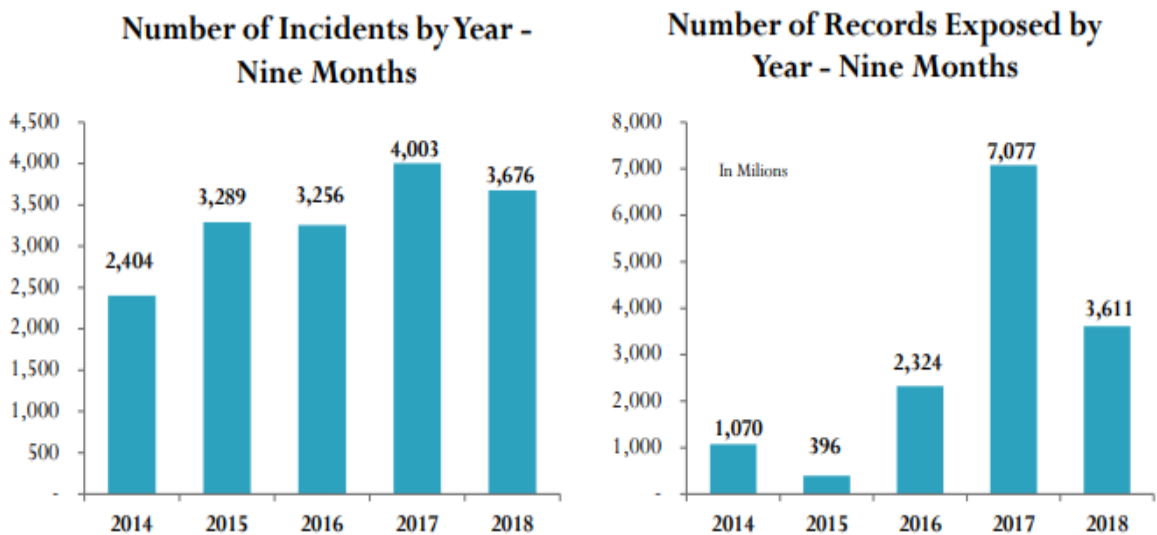

Figure 2: Data breaches in the first nine months of 2018 compared with previous years

Organizations with access to the most inclusive threat intelligence information bases offered and advanced search capabilities access to data via API and email alerting to help Enterprises make the right decisions in a very timely manner. This data is requested from numerous states and federal agencies within the United States, incorporating information through the freedom of Information Act (FOIA) requests.

The study presented in this work relies on 3,676 breaches that have been reported through September 30, 2018, revealing approximately 3.6 billion records. These records are restricted to occurrences fitting the criteria determined by the Open Security
Foundation. Furthermore, the data is balanced, and Redundancy is removed. Approximately every record separated from the database encompasses devoted fields for the data breach type, the source of a data breach, the affected countries and the textual description, and the affected data types. The diagrams in figure 2, figure 3, and figure 4 illustrates the no of total breaches in the year 2018 and its comparison with previous years, data breach sources, data breach by countries and affected data types accordingly. In this situation, the word unknown is being used to present data that is not available. This work focuses on DLP inside the organization, so data leakage from outside is not interesting.

Figure 2 shows that the number of reported breaches shows some improvements compared to the year 2017. The number of records dropped as organizations are focussing a lot on security and data leakage incidents. However, the decline from 2017 is only part of the story.2018 is on track to have the second most reported incidents and the third most exposed since 2005. Despite the number of breaches are less as compared to 2017. However, there is still a trend of overall breaches, creating more security concerns among organizations.

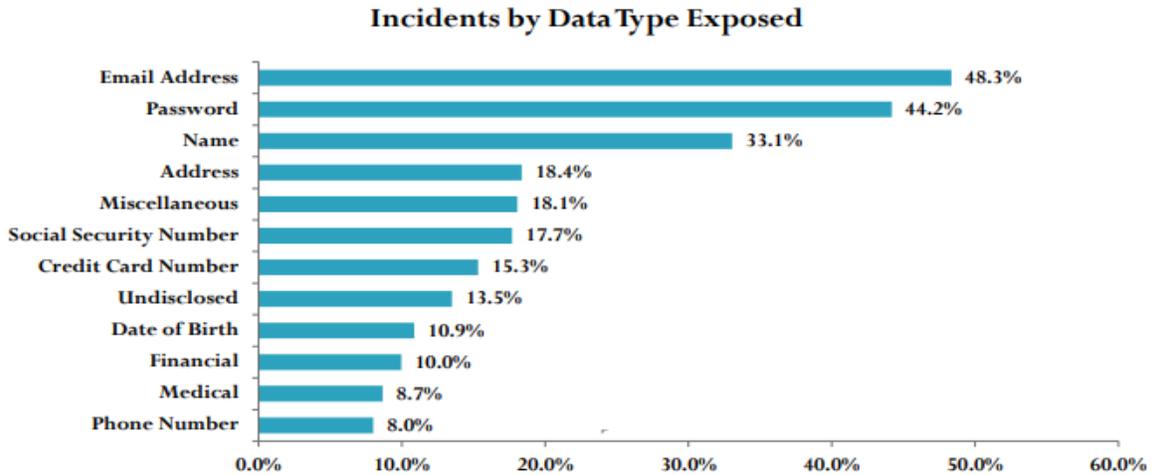

Figure 3: Datatypes affected by data breaches reported in 2018

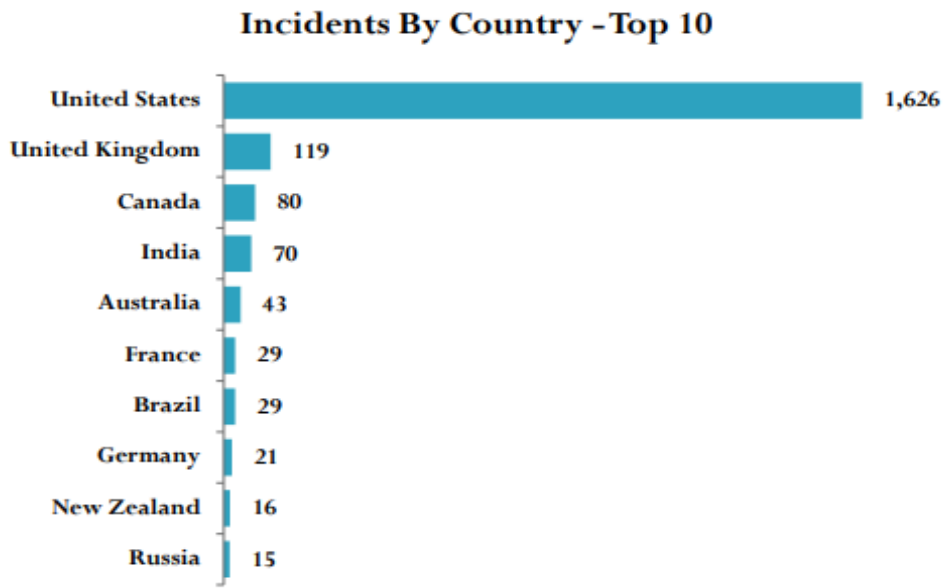

Figure 4: First Nine Months of Data breaches by Country in 2018

The data types affected by data breaches demonstrated in Figure 3 emphasize that the protection of email addresses, passwords, and names should be considered more precisely. Email address, user name, and user passwords are hacked from outside the organization; however, some social security numbers and credit card numbers are more vulnerable from inside the organization.

Figure 4 shows the top ten countries that accounted for the data breaches. It demonstrates that data leakage is not a problem for a single country. It is a worldwide problem that needs to be addressed carefully. Setting aside large breaches in China and Ukraine, the median number of records exposed is relatively high, indicating a larger number of breaches in those Countries.

## 3. Research Methodology

Usage of a DLP arrangement is an intricate endeavor that requires DLP architects to set up the exercises ahead of time. In this paper, our examination spotlights DLP execution using a combination of several techniques. It enlightens the way of its deployment inside the organization. The accompanying segments plot key contemplations for the DLP arrangement Process.

### 3.1 Proposed System Implementation

Usage of data transfer and sharing is expanding every day to be breached in several different ways. Nowadays, data security is a fundamental interest in large organizations. There is a huge amount of confidential data that resides inside the organization's premises. To prevent the data so that it will not be exposed to unauthorized entities, organizations focus a lot on the DLP solution's importance.

Data loss prevention is a complicated issue as it has no single effective solution. Authorities should explore DLP solutions according to the needs that best suit their organizational standards. To keep up the data's confidentiality and integrity, we suggested a data/information prevention system. Our system focuses on two main parameters, i-e, the state of the data, and its deployment procedure. Generally, data contains three different states from which there is a chance of loss or leakage, i-e data at rest (DAR), data in use (DIU), and data in motion (DIM). In this paper, DAR and DIU are out of context, so we will focus only on data in motion that travels from one network to another using HTTP, HTTPS, FTP, FTPS, and SMTP TCP/IP using SSL certificates and Deep traffic Inspection.

Figure 5 shows the flowchart diagram of our data protection system. As shown, there is a proxy server used in our DPS system i-e Squid Proxy server and C-Icecap server. Squid acts as an intermediary between a web user and a web server. The main function of the Squid proxy server is to break the connection between the web user and web server and transfer all the network traffic coming from
protocols like HTTP, HTTPS, FTP, FTPS, Webmail, SMTP, TCP/ IP, IMChat and pass that to ICAP server. C-ICAP is an extension of the ICAP server. It can be used with HTTP proxies for content filtering and adaptation. ICAP will then perform deep packet inspection (DPI) to look up for the sensitive information passing inside or outside the organization. After Packet inspection, if the system finds out that the data is encrypted, it will try to decrypt the data by the enterprise's keys. However, suppose the keys are not managed by the enterprise and provided to the DLP solution. In that case, our DLP system will automatically block user requests to proceed further.

In another case system will check the content being uploaded in combination with the techniques that we used in our system i-e
- ❖ Keywords and regular expressions
- ❖ Digital fingerprints
- ❖ Data tagging
- ❖ Content checking based on the use of Machine Learning Algorithms.

Along with using Keywords and Regular expressions, digital fingerprints are mostly used for digital signatures and for finding text fragments. Data tagging is also an approach to handle big data volumes. However, information creators or DLP Engineers should choose the tags. Moreover, to improve data loss/leakage's overall security, we implement more advanced content checkers based on machine learning. Instead of depending on specific dirty word lists from the database, we build data-driven solutions that automatically predict the word associated with that content. Still, care must be taken as it depends upon how we trained the system and how many data set we used for its learning process.

Taking associate actions on the DLP issues is the final objective of the DLP Program. As far as the actions are required, if our system found any Sensitive Information going outside the organization, it will automatically take necessary actions as defined by the organization rules and policies. Several possible actions

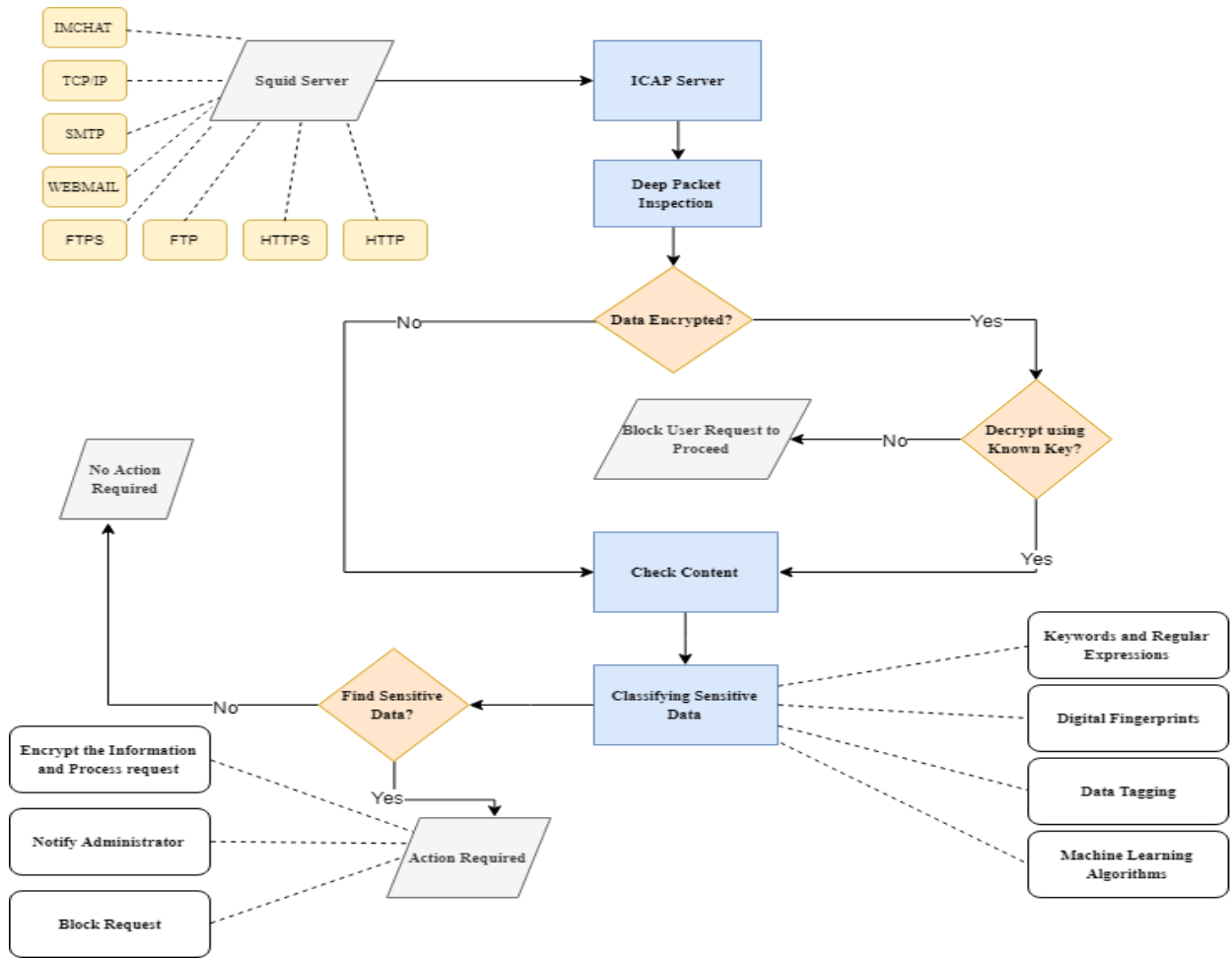

Figure 5:Flowchart for Data Protection System showing how DLP Solution is implemented

implemented in our system to protect the data leakage are

- ❖ Encryption of Information
- ❖ Notify Administrator
- ❖ Block User Request

Blocking the user request will alert the threat actor, and it will come up with two scenarios.

1. An attacker may attempt an alternate measure to send the data
2. DLP Engineer will not be able to detect the threat actor.

To prevent these falls out scenarios in our DLP solution, we encrypt the data using encryption algorithms and pass the user request to the webserver. It ensures that some shape of the information is sent back to an attacker is of no use. The eagerness to do this is to protect the data but not disappointing the attacker. The attacker attempts to leak the data that will be useful for us to improve the DLP solution further. Also, our system will notify the administrator about the user and his MAC Address. They tried to breach the organization's security.

## 3.2 Proposed System Deployment

To Preventing Data leakage/loss within the organization, the DLP solution might be deployed on three levels 1) Individual User Level 2) Domain Level 3) Gateway Level.

- ❖ Individual User Level: Some organizations hire their employees and allow them to work remotely from anywhere for their ease. In this scenario, their devices should be connected to their office device through an RSA Token.
- ❖ Domain-Level: Some organizations restrict their employees to use only the company domain to communicate with other
networks outside the world and prevent data leakage within the organization. Despite that, there are some special cases in which departments allow permission to some special users who can send data outside the network, which creates a chance of data leakage.
- ❖ Gateway Level: Gateway computer is a system that routes all the web traffic from one network to another. When an employee of a company sends anything outside the network, it passes through the standard gateway. In this way, data can easily be monitored and protected.

To inspecting the data that is going through the network, a DLP solution should be deployed in

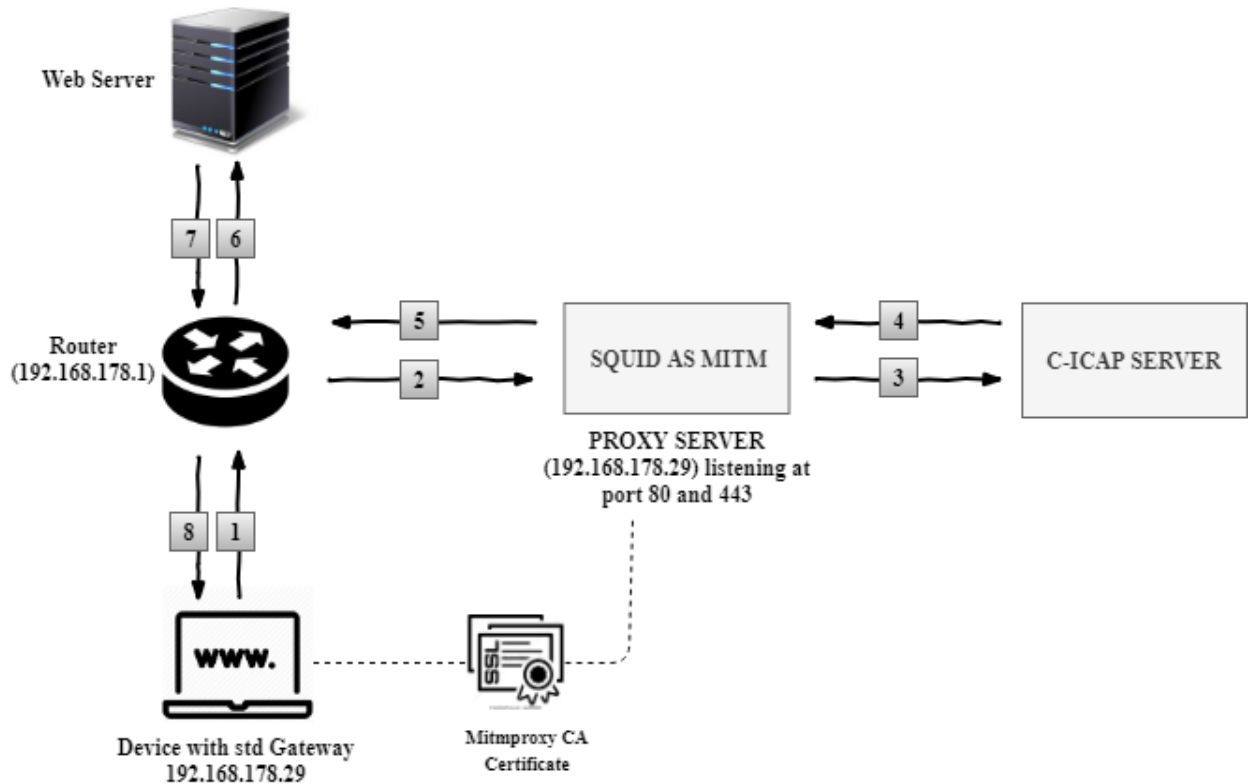

Figure 6: Logical Diagram showing how DLP solution is deployed within the organization

The platform has complete access to it, so we deployed our data protection system on the gateway level to protect confidential information.

Figure 6 shows a logical diagram of how our DLP solution is deployed inside the organizations. A user device with the standard gateway is the same as the device on which the squid server is running. When a user tries to send any information through the network, it passes from the gateway. Instead of going directly from gateway to web server, traffic will be passed through the device on which squid server and C-ICAP server are running to monitor and protect the confidential information. SSL Certificates are also generated by the Squid server, which is managed by using group policies on a domain controller. Instead of manually adding certificates in every device inside an organization using an enterprise domain, it will be automatically added by running scripts on the gateway level.

## 3. Conclusion and Future Work

Data leakage is a key factor that damages a company's reputation. Most of the data and Information are leaked from internal sources. This paper provides a recent survey report on data and information leakage incidents in the year 2018. The analysis of data breaches reported in 2018 dropped as compared to the year 2017. This paper also explains why there is a need for a DLP solution, how data is moving through intermediate channels before reaching its destination, and what necessary actions should be taken to protect sensitive data.

Our DLP solution is the combination of previous methodologies with advanced methodologies to ensure that no sensitive data goes outside the organization by any means to increase the confidentiality and integrity of an enterprise. This proposed system can block, notify, and encrypt the information while passing through the gateway. Also, our system notifies the administrator about the user and his MAC address. They attempt to breach the data to enquire against him.

Future efforts can be made in implementing the DLP solution in cloud computing as most companies are moving their data to the cloud, which leads to security and compliance concerns.

## Data Availability

The data used to support the findings of this study are available from the first author upon request.

## Conflicts of Interest

We declare that we do not have any commercial or associative interest that represents conflicts of interest connected with the work submitted.

## Acknowledgments

The authors would like to acknowledge the support provided by the National Key R&D Program of China (No.2018YFC1604000/2018YFC1604002) and the Natural Science Foundation of Hubei province (No.2017CFB663).